\documentclass[prb,twocolumn,aps,showpacs]{revtex4}
\usepackage{graphicx}
\usepackage{subfigure}
\begin{document}

\title{Carbon Nanotubes as Nanoelectromechanical Systems}

\author{S.~Sapmaz, Ya.~M.~Blanter, L.~Gurevich, and H.~S.~J.~van der
Zant}
\affiliation{
Department of NanoScience and DIMES, Delft University of
Technology, Lorentzweg 1, 2628 CJ Delft, The Netherlands}
\date{\today}
\begin{abstract}
We theoretically study the interplay between electrical and mechanical
properties of suspended, doubly clamped carbon nanotubes in which
charging effects dominate. In this geometry, the capacitance between
the nanotube and the gate(s) depends on the distance between
them. This dependence modifies the usual Coulomb models and we show
that it needs to be incorporated to capture the physics of the problem
correctly. We find that the tube position changes in discrete steps
every time an electron tunnels onto it. Edges of Coulomb diamonds
acquire a (small) curvature. We also show that bistability in the tube
position occurs and that tunneling of an electron onto the tube
drastically modifies the quantized eigenmodes of the tube.
Experimental verification of these predictions is possible in
suspended tubes of sub-micron length.
\end{abstract}

\maketitle

\section{Introduction}

Nanoelectromechanical systems (NEMS) convert electrical current into
mechanical motion on a nanoscale and vice versa. They can be viewed as
the successors~\cite{Roukes1} of microelectromechanical-devices (MEMS)
which operate at a micron scale and which are found in commercial
applications. Improved performance is expected from NEM-devices due to
their small sizes, and higher eigenfrequencies. M(N)EMS have already
been used for high-precision measurements of force \cite{Att},
electric charge  \cite{Cleland}, the thermal conductance quantum
\cite{Roukes2}, and the Casimir force \cite{Casimir}.
From a fundamental point of view, NEM-physics is an unexplored
field in which new phenomena are likely to be found. Examples include
tunneling through moving barriers \cite{Roukes98}, additional sources
of noise \cite{Levitov}, and shuttling mechanism for transport
\cite{Goth,Erbe,C60}.
\vspace{1.0cm}

\begin{figure}[ht]
\includegraphics[angle=0,width=6.3cm]{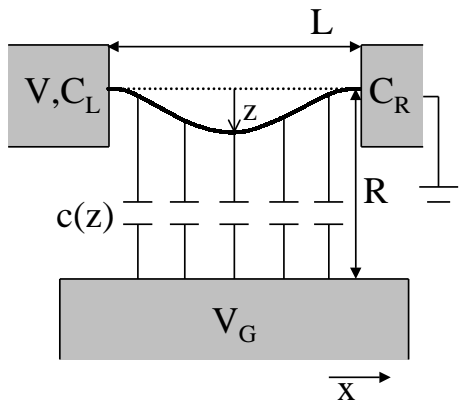}
\caption{\label{Fig1} A schematic drawing of a suspended nanotube
capacitively coupled to a gate and clamped on both sides to metal pads
that serve as tunnel contacts. A voltage $V$ is applied to the left pad.}
\end{figure}

Studies with NEMS have mostly been performed in devices made with
silicon technology. Carbon nanotubes provide an interesting
alternative because of their superior mechanical properties. They have
already been implemented as nanotweezers~\cite{kim,akita}, as switches
in a random access memory device~\cite{rueckes}, or as nanoscale
actuators~\cite{sensors}. In addition, recent theoretical
calculations show that carbon nanotubes can also be used as
nanoelectromechanical switches~\cite{dequesnes,kinaret} or as
gigahertz oscillators~\cite{legoas}.

In this paper, we study theoretically nanoelectromechanical effects in
doubly-clamped suspended carbon nanotubes. Doubly-clamped suspended
single- and multi-wall carbon nanotubes have been previously
fabricated, and their
transport~\cite{Tombler,Cobden,Williams,Franklin}, acoustoelectric
\cite{freq}, thermal \cite{PKim}, and elastic \cite{GTKim} properties
have been measured. We consider a single-wall carbon nanotube (SWNT)
in which Coulomb-blockade effects dominate transport, and demonstrate
that a gate manipulates the tube in an effective way. The
applied gate voltage bends the tube, changes the stress and thus
influences the electric and mechanical properties. 

This paper is organized as follows: The next Section describes the
model with inclusion of the influence of initial stress and thermal
fluctuations. We concentrate on the case where the junction
capacitances are zero so that analytical expressions are
obtained. Section \ref{Couleff} describes the influence of
nanoelectromechanical effects on Coulomb blockade and shows that
intrinsic bistability occurs when the tube is strained. Section
\ref{Eigenmodes} discusses the eigenmodes and the
influence on the initial strain on them. In Section \ref{relax}
junction capacitances are no longer neglected and we also show the
effect of a non-uniform charge distribution. We end with some remarks
on the limitations of our model.

\section{Displacement, stress, and energy} \label{dse}

\subsection{Equilibrium position}

We consider a SWNT (modeled as a rod of length $L$ along the
$x$-axis), freely suspended between source and drain electrodes, in
the vicinity of a gate (see Fig.~1). The nanotube is attached to the
electrodes via tunneling contacts. An electrostatic force (gate
voltage) bends the tube; the deviation from a straight line is denoted
by $z(x)$ with $0 < x < L$. The elastic energy of the bent tube is
\cite{LL}
\begin{eqnarray} \label{elastic}
W_{el} [z(x)] & = & \int_0^L dx \left\{ \frac{EI}{2} z''^2
\right. \nonumber \\
& + & \left. \left[
\frac{T_0}{2} + \frac{ES}{8L} \int_0^L z'^2 dx \right] z'^2 \right\}
,
\end{eqnarray}
where $E$, $I=\pi r^4/4$, and $S = \pi r^2$ are the elastic modulus,
the inertia moment and the cross-section, respectively. Here, $r$ is
the (external) radius of the tube. The first term in
Eq. (\ref{elastic}) is the energy of an unstressed bent rod; the two
other terms describe the effect of the stress force $\tilde T = T_0 +
T$. Here $T_0$ is the residual stress which may result {\em e.g.} from
the fabrication, and the induced stress $T$ is due to the elongation
of the tube caused by the gate voltage,
\begin{equation} \label{stress1}
T = \frac{ES}{2L} \int_0^L z'^2 dx.
\end{equation}

To write down the electrostatic energy, we denote the
capacitances of the barriers connecting the nanotube with the source
and drain as $C_L$ and $C_R$, respectively (see Fig.~1). The
capacitance to the gate per unit length is $c(z)$. Approximating the
gate by an infinite plane at a distance $R$ from the nanotube, we
obtain
\begin{equation} \label{capac1}
c (z) = \frac{1}{2 \ln \frac{2(R-z)}{r}} \approx \frac{1}{2 \ln
\frac{2R}{r}} + \frac{z(x)}{2R \ln^2 \frac{2R}{r}},
\end{equation}
where the Taylor expansion restricts validity to $z \ll R$. In this
limit van der Waals forces between the nanotube and the substrate can 
be neglected. The electrostatic energy of the system reads
\begin{eqnarray} \label{elst}
& & W_{est} [z(x)] = \frac{(ne)^2 -2ne(C_LV + C_GV_G)}{2(C_L+C_R+C_G)}
\\
& & - \frac{C_L(C_R+C_G)V^2 + C_G(C_L+C_R)V_G^2 -
2C_LC_GVV_G}{2(C_L+C_R+C_G)}, \nonumber
\end{eqnarray}
where $V$ and $V_G$ are the potentials of the source and the gate
respectively (the drain potential is set to zero), $ne$ is the
(quantized) excess charge on the nanotube, and for a uniform charge
distribution the capacitance to the gate equals
\begin{displaymath}
C_G = \int_0^L c[z(x)] dx.
\end{displaymath}
Note, that the last term in Eq. (\ref{elst}) depends on the tube
displacement and thus on the number of
electrons. Therefore, it can not be omitted as in the standard
Coulomb blockade treatment that replaces this term by a constant making
$W_{est}$ a periodic function of gate voltage.

In the following, we concentrate on the analytically tractable case
$C_L, C_R = 0$. The general case is considered in Section
\ref{relax}. For a moment, we also assume
$T_0 = 0$. In this situation, the expression for the electrostatic
energy simplifies,
\begin{eqnarray} \label{elst1}
& & W_{est} [z(x)] = \frac{(ne)^2}{2C_G[z]} - neV_G \nonumber \\
& & \approx \frac{(ne)^2 \ln \frac{2R}{r}}{L} - \frac{(ne)^2}{L^2R}
\int_0^L z(x)dx - neV_G .
\end{eqnarray}
Minimizing the energy,
\begin{displaymath}
W_n [z(x)] = W_{el} [z(x)] + W_{est} [z(x)],
\end{displaymath}
with respect to $z$, one finds the equation determining the tube
position~\cite{LL},
\begin{eqnarray} \label{eqs1}
IEz'''' - T z'' = K_0 \equiv \frac{(ne)^2}{L^2R},
\end{eqnarray}
where $K_0$ is the electrostatic force per unit length, which we
approximate by a constant. Higher-order terms are small for $z \ll
R$. To solve Eq.~(\ref{eqs1}) we have to assume that the stress force
$T$ is constant, and find it later from the self-consistent condition
(Eq.~(\ref{stress1})).

The solution of Eq. (\ref{eqs1}) with the appropriate boundary
conditions (for the doubly-clamped rod $z(0) = z(L) = z'(0) = z'(L) =
0$) has the form
\begin{eqnarray} \label{solution1}
z_n (x) & = & \frac{K_0L}{2T \xi} \left[ \frac{\sinh \xi L}{\cosh \xi
L - 1}(\cosh \xi x - 1) - \sinh \xi x \right. \nonumber \\
& + & \left. \xi x - \xi \frac{x^2}{L} \right], \ \ \ \ \ \xi =
\sqrt{\frac{T}{EI}}.
\end{eqnarray}
Substituting this into Eq. (\ref{stress1}), a relation between the
stress $T$ and
the external force $K_0$ is obtained. In the limiting cases, it reads
\begin{eqnarray} \label{stress2}
T = \left\{ \begin{array}{lr}
K_0^2L^6S/(60480EI^2), & T \ll EI/L^2, \\
(ES/24)^{1/3}(K_0L)^{2/3}, & T \gg EI/L^2.
\end{array} \right.
\end{eqnarray}
The first line corresponds to weak bending of the tube: The
energy associated with the bending exceeds the energy of the
stress. Generally, it is realized for $z \lesssim r$. The second line
describes strong bending, when the tube displacement is large ($ r < z
\ll R, L$).

For the displacement of the tube center $z_n^{max} = z_n (L/2)$ we
find
\begin{eqnarray} \label{displace}
\begin{array}{lrl}
z_n^{max} = 0.003 \frac{(ne)^2 L^2}{Er^4 R}, & \ \ T \ll
\frac{EI}{L^2} & \ \ \left( n \ll \frac{Er^5R}{e^2L^2} \right); \\
z_n^{max} = 0.24 \frac{(ne)^{2/3} L^{2/3}}{E^{1/3}r^{2/3} R^{1/3}}, & \
\ T \gg \frac{EI}{L^2} & \ \ \left( n \gg \frac{Er^5R}{e^2L^2}
\right).
\end{array}
\end{eqnarray}
For a SWNT with $r = 0.65$ nm, $E = 1.25$~TPa,
$L = 500$ nm and $R = 100$ nm (to be referred to as the E-nanotube)
the crossover from weak to strong bending, $T \sim EI/L^2$, occurs
already at $n \sim 5 \div 10$. In the strong-bending
regime, the displacement of the E-nanotube is (in
nanometers) $z_n^{max} = 0.24 n^{2/3}$. Note that this regime is not
accessible with state-of the art silicon submicron devices, which are
always in the weak-bending limit.
\begin{figure}[ht]
\includegraphics[angle=0,width=8.cm]{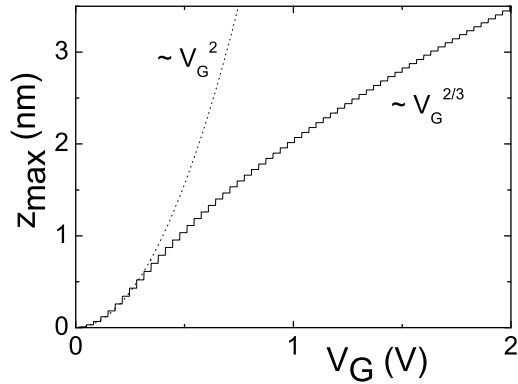}
\caption{\label{Fig2} Calculated displacement as a function of gate
voltage for the E-nanotube: $r = 0.65$ nm, $E = 1.25$~TPa, $L
= 500$ nm and $R = 100$ nm. At $V_G \approx 0.5V$, there is a
crossover from weak bending with a $V_G^2$-dependence to strong
bending with a $V_G^{2/3}$ dependence. }
\end{figure}

\subsection{Charge and energy}

For comparison with experiments, we have to relate the charge
$ne$ to the gate voltage by minimizing the energy. The expression for
the energy (elastic plus electrostatic) of the tube at equilibrium in
the limiting cases reads
\begin{eqnarray} \label{energy5}
W_n^{eq} & \equiv & W_{st} - \delta W = \frac{(ne)^2}{L} \ln \frac{2R}{r}
- neV_G \\
& - & \left\{ \begin{array}{lr} 0.0009(ne)^4L/(Er^4R^2), \ \ \ \ \
\  T \ll EI/L^2; \\
0.08(ne)^{8/3}/(Er^2R^4L)^{1/3}, \ \ \  T \gg EI/L^2.
\nonumber \end{array} \right.
\end{eqnarray}
The first two terms represent the electrostatic energy of a straight
tube, and the third one is due to the elastic degrees of freedom
(stress, bending, and change of $C_G$ due to displacement). This
nonlinear, {\em nanomechanical term} is typically a small correction:
For the E-nanotube it becomes of the same order as $W_{st}$ if $n \sim
3000$ in which case Eq. (\ref{capac1}) is not valid anymore.
The negative sign of the nanomechanical contribution is easily
understood: As the gate voltage changes, the movable tube adjusts not
only its charge, but also its position, which leads to a
lower energy as compared to the fixed-position system.

The value of $n$ which minimizes the energy is
\begin{displaymath}
n = Int \left( \frac{V_G L}{2e \ln (2R/r)} + \frac{1}{2} + \delta n
\right),
\end{displaymath}
with $Int$ denoting the integer part of the expression. The small
correction $\delta n$ in the strong-bending regime is proportional to
$V_G^{5/3}$. Thus, the tube displacement $z_{max}$ changes in discrete
steps when $V_G$ is varied as shown in Fig.~2.
The envelope is proportional to $V_G^2$
(weak bending) or $V_G^{2/3}$ (strong bending). In the absence of
charging effects and tension, the displacement is given by the dashed
line as previously found in simulations of
Ref. \onlinecite{dequesnes}.  

\subsection{Thermal fluctuations}

The preceding considerations are restricted to the case of zero
temperature. To understand the role of the temperature, we now
evaluate the effect of thermal fluctuations on the equilibrium
position of the tube.

The variance of the position of the tube center at a given
charge $n$ can be generally represented as a functional integral,
\begin{eqnarray} \label{funcint}
& & \mbox{var}\ z_n \equiv \left\langle \left[ z(L/2) - z_n (L/2)
\right]^2 \right\rangle \nonumber \\
& = & \left. \frac{\partial^2}{\partial J^2} \int Dz(x)
\exp \left[-W_n[z]/k_B\Theta + J z(L/2) \right]\right\vert_{J=0}
\nonumber \\
& \times & \left[ \int Dz(x)
\exp \left( -W_n[z]/k_B\Theta \right) \right]^{-1},
\end{eqnarray}
where $\Theta$ is the temperature. Except for $n=0$, the functional
integral in Eq. (\ref{funcint}) is not Gaussian and has to be
linearized around the equilibrium solution $z_n(x)$, Eq.
(\ref{solution1}). The remaining Gaussian integral can be
calculated, and we arrive at
\begin{equation} \label{fluct1}
\mbox{var} \ z_n = k_B \Theta \zeta(L/2),
\end{equation}
where $\zeta(x)$ solves the equation
\begin{eqnarray} \label{fluct2}
EI\zeta'''' - \frac{ES}{2L} \int z_n'^2 dx \ \zeta'' - \frac{ES}{L}
z_n'' \int \zeta' z_n' dx \nonumber \\
= \delta (x - L/2).
\end{eqnarray}

In the two limiting cases of weak and strong bending, the solution
of Eq. (\ref{fluct2}) yields
\begin{eqnarray} \label{fluct3} \mbox{var}
\ z_n = \left\{ \begin{array}{lr} k_B\Theta L^3/192 EI, & n=0 \\ k_B
\Theta L/8T, & n \gg Er^5 R/e^2 L^2 \end{array} \right. \ ,
\end{eqnarray}
where the stress $T$ is still given by the lower line of
Eq. (\ref{stress2}).
Thus, the fluctuations in the tube position are expected to grow
linearly with temperature. However, their magnitude is small.
For the E-nanotube, at 100K the fluctuations in the
$n=0$ state are of the order of $0.1$ nanometer, and at least an
order of magnitude less in the strong-bending regime.

In the calculations, we have assumed that the charge $ne$ is a fixed
quantity. Close to the degeneracy points $W_n^{eq} = W_{n+1}^{eq}$
thermal fluctuations may induce switching between the states with
charges $ne$ and $(n+1)e$, in which case Eq. (\ref{fluct3}) is no
longer valid. However, the range of voltages where switching is
important, is narrow. 

\section{Coulomb effects and bistability} \label{Couleff}

\subsection{Coulomb blockade}

Since the nanotube is attached to the electrodes by tunneling contacts,
it is in the Coulomb blockade regime. We define the energy to add the
$n$th electron to the tube as $S_n = W_{n}- W_{n-1}$. Then, if the
nanotube contains $n>0$ electrons, the conditions that current can not
flow (is Coulomb blocked) are $S_n < 0, eV < S_{n+1}$.  In
quantum dots, $S_n$ depends linearly on the bias $V$ and gate
$V_g$ voltages. Thus, in the $V_G - V$ plane regions with zero
current are confined within Coulomb diamonds, that are identical
diamond-shape structures repeating along the $V_G$--axis.

In a suspended carbon nanotube, in addition to the purely Coulomb
energy, we also have the nanomechanical corrections. Generally, these
corrections make the relations between $V$ and $V_G$, which describe
the boundaries of Coulomb blockade regions, non-linear. Consequently,
the Coulomb ``diamonds'' in suspended nanotubes are not diamonds any
more, but instead have a curvilinear shape (with the exception of the
case $C_L = C_R = 0$). Their size is also not the same and decreases
with $\vert V_G \vert$. Thus, the mechanical degrees of freedom {\em
affect} the Coulomb blockade diamonds. However, since these effects
originate from the nanomechanical term which is typically a small
correction, its influence on Coulomb diamonds is small as well. For
the E-nanotube, these effects do not exceed several percents for
typical gate voltages. 

\subsection{Two-gate setup and bistability}

To demonstrate that the nanomechanical effects can not generally be
omitted, we consider a suspended tube symmetrically placed in between
two gates and show below that
bistability in the tube position occurs~\cite{Armour1}.

Fig.~1 again presents the schematic setup, but the suspended tube is
placed between two gates, labeled up (U) and down (D). Since up and
down capacitances are connected in parallel, their 
sum $C_G = C_U + C_D$ matters. Assuming that the distance of the
straight tube to both gates is the same, we write
\begin{equation} \label{capac2}
C_{U,D} = \int_0^L \frac{dx}{2 \ln \frac{2(R \mp z)}{r}},
\end{equation}
Expanding this for $z \ll R$ and calculating the electrostatic force,
we arrive at an equation similar to Eq. (\ref{eqs1}), with a
constant force $K_0$ that is replaced by $\gamma z$, where
\begin{displaymath}
\gamma = \frac{(ne)^2 (\ln 2R/r + 2)}{2L^2R^2 \ln 2R/r}.
\end{displaymath}
We now solve this equation in the strong-bending regime. For this
purpose \cite{LL} we disregard the term $IEz''''$, and use the
boundary conditions $z(0) = z(L) = 0$. Multiple solutions emerge; the
ones with the lowest energy are
\begin{equation} \label{multistab}
z = \pm \frac{2L^2}{\pi^2} \sqrt{\frac{\gamma}{ES}} \sin \frac{\pi
x}{L}.
\end{equation}
Thus, the tube in the strong-bending regime can oscillate between the
two symmetric positions. This creates a basis for observation
of quantum effects, as discussed in Ref. \onlinecite{Armour1}. We
emphasize once again, that within this model, the multi-stability is
due to the charging of the tube in combination with the
non-linearity.

\section{Eigenmodes} \label{Eigenmodes}

The eigenfrequency of a particular eigenmode is an important directly
measurable\cite{freq} property. In future experiments on suspended
tubes we expect that the eigenmodes influence tunneling
("phonon-assisted tunneling") in a similar way as observed for a
single $C_{60}$ molecule~\cite{C60}. Below, we demonstrate that the
effect of the electrostatic interactions on the elastic properties
(specifically, eigenfrequencies) is strong and changes the behavior
qualitatively.

To find the eigenmodes, we apply a gate voltage with a large dc
(single gate) and a small ac component. The displacement $z(x,t)$ is
time-dependent, which provides an external force $-\rho S
\ddot{z}$ to Eq. (\ref{eqs1}), where $\rho$ equals $1.35$
g/cm$^3$. Eq. (\ref{eqs1}) must be solved first with a constant
stress, and then the stress is found self-consistently. The tube
displacement has a small ac component $\delta z$ on top of a large
static one. The self-consistency procedure is essentially the same and
again leads to Eq. (\ref{stress2}). Thus, the dc component of the gate
voltage determines the stress $T$ and it therefore controls the
eigenmodes.

The frequencies of the (transverse) eigenmodes are found from the
requirement that the equation
\begin{equation} \label{modes1}
IE \delta z'''' - T \delta z''- \rho S \omega^2 \delta z = 0
\end{equation}
with the boundary condition $\delta z(0) =
\delta z(L) = \delta z'(0) = \delta z'(L) = 0$ has a non-zero solution.
This yields the following equation for the frequency $\omega$,
\begin{eqnarray} \label{modes2}
& & \cosh y_1 \cos y_2 - \frac{1}{2} \frac{y_1^2 - y_2^2}{y_1y_2} \sinh
y_1 \sin y_2 = 1, \\
& & y_{1,2} = \frac{L}{\sqrt{2}} \left( \sqrt{\xi^4 + 4 \lambda^2} \pm \xi^2
\right)^{1/2}, \ \ \ \lambda = \sqrt{\frac{\rho S}{EI}}
\omega. \nonumber
\end{eqnarray}
In the following, we restrict ourselves to the fundamental (lowest
frequency) eigenmode $\omega_0$. In the limiting cases, the
solutions of Eq. (\ref{modes2}) are
\begin{eqnarray} \label{modes3}
\omega_0 = \sqrt{\frac{EI}{\rho S}} \left\{ \begin{array}{lr}
22.38L^{-2} + 0.28\xi^2, & \xi L \ll 1; \\
\pi \xi L^{-1} + 2\pi L^{-2}, & \xi L \gg 1.
\end{array} \right.
\end{eqnarray}
The second terms on the rhs represent small corrections to the
first ones.

The frequency dependence $\omega_0 \propto L^{-2}$ is associated with a
loose string, while $\omega_0 \propto L^{-1}$ means that the
string is tied like in a guitar. Our results show that the behavior of
the tube crosses over from ``loose'' to ``tied'' as $V_G$
increases. For the fundamental mode, the crossover occurs at $\xi L
\sim 1$, corresponding to the crossover from weak to strong bending.
The middle curve in Fig.~3 shows the frequency of the fundamental
mode as a function of gate voltage (zero residual stress). The arrow
denotes the cross-over from weak to strong bending.

The gate voltage dependence of the frequency is a stepwise
function, as shown in the inset of Fig.~3. Steps occur
whenever an additional electron tunnels onto the tube. For the
E-nanotube, their height is $\sim$5~MHz, which is measurable.
Note, that the present submicron silicon devices are always
in the weak-bending regime so that corrections due to the second term
in Eq.~(\ref{modes3}) are too small to be measured.
Furthermore, one should realize that frequency quantization is only
observable if the frequency itself is greater than the inverse
tunneling time for
electrons.

We now consider the effect of a residual stress ($T_0 \neq
0$). First, we obtain the stress by solving Eqs. (\ref{stress1}),
(\ref{eqs1}) (in the latter, $T$ is replaced by $T+T_0$). In
particular, for a negative stress $T+T_0 < 0$, $T_0 \sim
-EI/L^2$, Eq. (\ref{stress1}) acquires several solutions.
This signals {\em Euler instability}: the tube bends in the
absence of an external force.
\vspace{1.8cm}

\begin{figure}[ht]
\includegraphics[angle=0,width=7.cm]{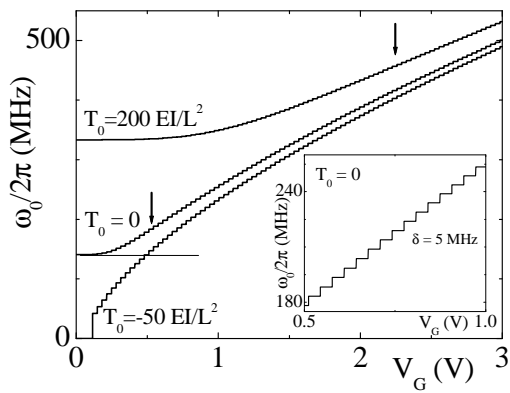}
\caption{\label{Fig4} Gate voltage dependence of the frequency
$\omega_0$ of the fundamental mode for three different values of the
residual stress. Numbers are taken for the E-nanotube (see
Fig.~2). The fundamental mode of an unstressed tube is 140 MHz (thin
horizontal line). The inset is an enlargement of the $T_0=0$ curve of
the main figure showing step-wise increases of $\omega_0$ whenever an
additional electron tunnels onto the tube.}
\end{figure}

If the residual stress is large, $T_0 \gg EI/L^2$, the tube always
acts like a tied string (upper curve in Fig.~3). The frequency
depends weakly on $V_G$ for low voltages, and above $T \sim T_0$
(denoted with the arrow) grows with an envelope $\propto
V_G^{2/3}$. For negative $T_0$ the picture is
qualitatively different (lower curve in Fig.~3). Whereas for large
gate voltages the envelope is still proportional to
$V_G^{2/3}$, the frequency dives below the value for an unstressed
tube ($22.38(EI/\rho S)^{1/2}L^{-2}$, represented by the thin solid
line in Fig.~3), when the overall stress becomes
negative. It further drops to zero at the Euler instability
threshold.

The qualitative difference between the various regimes means that by
measuring the gate voltage dependence of $\omega_0$
one can determine the sign of $T_0$ and get a quantitative estimate.
On the other side, the gate effect can be used to tune the
eigenfrequencies. We also mention that in the absence of charging
effects, the steps vanish but the overall shape of the curves in
Fig.~3 remains the same.

\section{Relaxing the approximations} \label{relax}

While considering equilibrium displacement and eigenmodes of the
nanotube, we made a number of simplifying approximations. In this
Section, we consider two of them --- disregarding the capacitances
$C_{L,R}$ and uniform distribution of the charge --- and show that
relaxing these approximations affects the above results
quantitatively, but not qualitatively.

In this Section, we consider the case of zero residual stress $T_0 =
0$.

\subsection{Finite capacitances to the leads}

\begin{figure}[ht]
\includegraphics[angle=0,width=7.cm]{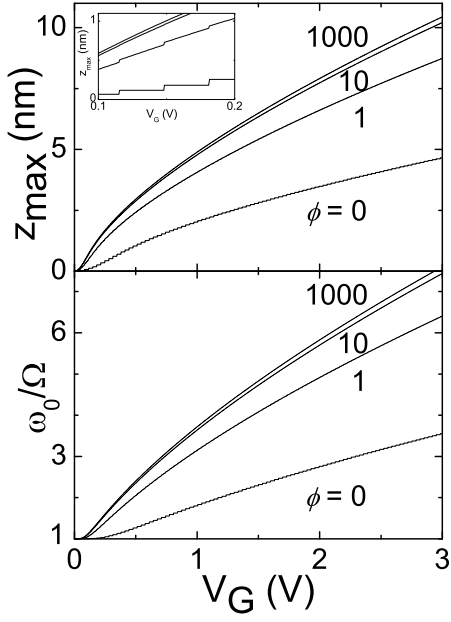}
\caption{Above: Displacement as a function of gate voltage for the
E-nanotube with finite capacitances to the leads. The four curves
correspond to different values of the parameter $\phi$, defined as
$C_L=C_R=\phi C_G$. The inset is an enlargement of the main
figure. Below: The frequency of the fundamental mode normalized to the
fundamental frequency of an unstressed tube $\Omega=(22.38 L^{-2}
(EI/\rho S)^{1/2} = 141$ MHz for the same parameters as above. }
\label{disp01} 
\end{figure}

We now relax the limitation $C_L, C_R = 0$. For the general case,
Eq. (\ref{eqs1}) still holds, however, the force $K_0$ must be
adjusted,
\begin{eqnarray}\label{forcegen}
K_0 & = & \frac{1}{L^2R}
\frac{C_0^2}{(C_0+C_R+C_L)^2} \nonumber \\
& \times & \left[ ne+(C_L+C_R)V_G-C_LV \right]^2,
\end{eqnarray}
where $C_0=L/(2\ln 2R/r)$ is the capacitance of the straight nanotube
to the gate. The results of the numerical solutions for the
displacement and the frequency of the fundamental mode are plotted in
Fig. \ref{disp01}. For simplicity we have taken $C_L=C_R=\phi
C_G$; the four curves correspond to different values of the
parameter $\phi$. The curves with $\phi = 0$ are the same as the
ones in Figs.~\ref{Fig2},~\ref{Fig4}.

The plots demonstrate that the qualitative picture remains the
same if we include finite capacitances to the leads. The steps
observed for $\phi=0$ become skewed with the increase of $C_L$ and
$C_R$ (see inset of Fig.~4). At a certain $\phi$ they disappear.
For $\phi>10$ the plots are, on the scale presented, the same.

\subsection{Non-uniform charge distribution}

Above, we have assumed a uniform charge distribution along the
nanotube. Rather than trying to analyze the effect in general, we
consider the opposite situation when the excess charge is concentrated
at one point (to be more precise, in a concise region of the tube
radius $r$), which may represent, for instance, a pinning center.
This center is placed in the middle of the nanotube.
Though we believe that the charge distribution
in suspended nanotubes is closer to uniform, this
situation applies to a suspended quantum dot as realized
recently~\cite{Munich}.

\begin{figure}[ht]
\includegraphics[angle=0,width=7.cm]{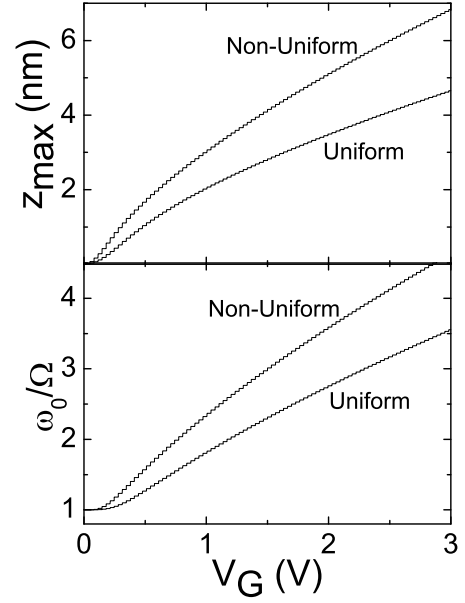}
\caption{Displacement of the center of the E-nanotube (above) and the
frequency of the fundamental eigenmode (below) for uniform
charge distribution and for the case that the charge is concentrated
at one point. $\Omega$ is the fundamental frequency of an unstressed
tube.} \label{nonuniform}
\end{figure}

The gate-charge capacitance $C_G$ in this geometry is
\begin{equation}\label{Capicatance gate charge}
C=\frac{1}{\frac{1}{r}-\frac{1}{2R}},
\end{equation}
and we proceed to obtain the equations of motion,
\begin{equation}\label{point1}
IEz''''-Tz''= F \delta \left( x-\frac{L}{2} \right), \ \ \ F \equiv
\frac{(ne)^2}{4R^2},
\end{equation}
where we again set $C_L = C_R = 0$. 

The solution with the same boundary conditions as previously, $z(0) =
z(L) = z'(0) = z'(L) =0$, and with $z$, $z'$, and $z''$ all continuous
at $x=L/2$, is
\begin{eqnarray}\label{solutionpoint}
z(x) & = & \frac{F}{2EI \xi^3} \\
& \times & \left\{ \tanh \frac{\xi L}{2}
[\cosh \xi x -1]-\sinh \xi x  +\xi x \right\} \nonumber
\end{eqnarray}
for $0 < x < L/2$. For $L/2 < x < L$ the coordinate $x$ should
be replaced by $(L-x)$ because $z(x) = z(L-x)$. As before, $\xi =
(T/EI)^{1/2}$ and
Eq. (\ref{stress1}) is used to obtain the stress self-consistently,
\begin{eqnarray}\label{stress3}
T = \left\{ \begin{array}{lr} F^2L^4S/(30720 EI^2), & T \ll EI/L^2, \\
(1/2)(ESF^2)^{1/3}, & T \gg EI/L^2. \end{array} \right.
\end{eqnarray}

Consider now the strong-bending regime and compare the results for the
stress $T_u$ for the uniform (lower line of Eq. (\ref{stress2}) and
$T_n$ for the concentrated (lower line of Eq. (\ref{stress3}) charge
distributions,
\begin{equation}\label{stress4}
T_n =T_u \left(\frac{\sqrt{3}L}{4R}\right)^{2/3}.
\end{equation}
For $L \gg R$ we formally have $T_n \gg T_u$. This means that for the
same gate voltage more stress is induced at the nanotube if the charge
is concentrated at one point. Also, the displacement of the tube is
greater in the concentrated case,
\begin{displaymath}
z^n_{max} = 0.87 \left( \frac{L}{R} \right)^{1/3} z^u_{max}.
\end{displaymath}
Thus, if the charge distribution is concentrated, NEMS are ``more
effective'' than for the uniform charge. For the E-nanotube the ratio
between non-uniform and uniform maximal displacement is $1.49$. The
difference between uniform and non-uniform charge distributions is 
illustrated in Fig.~\ref{nonuniform}.

\section{Discussion}

The presented model is simplified in many respects. Mechanical degrees
of freedom are introduced via classical theory of elasticity: The
nanotube (modeled by a rod) is considered as incompressible and
without internal structure. This is justified, since so far the theory
of elasticity has described all existing experiments on carbon
nanotubes well. For SWNTs it has been also supported by simulations
(Ref. \onlinecite{dequesnes}). Creation of defects in SWNT starts at
deformations of order of ten percents. For larger deformations (see
{\em e.g.} Ref. \onlinecite{Yakobson}) we expect strong deviations 
from the behavior we describe, but this typically lies outside our
applicability range $z \ll R$. We have neglected damping, which is
also expected to originate from the creation of the defects and to be
irrelevant in this range. We also disregarded quantum effects
(cotunneling and finite spacing of quantum levels of electrons in the
tube). These issues need to be clarified for a detailed comparison
with the experimental data, and will be a subject for future research.

Our main result is that the nanotube can be manipulated by the gate
voltage, which determines its deformation and stress, and modifies the
eigenmodes. Though the eigenmodes of nanotube ropes have been
measured in Ref. \onlinecite{freq} three years ago, the {\em strain
dependence} of the eigenmodes was only recently reported in
Ref. \onlinecite{purcell} which was published after this manuscript
had been submitted for publication. Ref. \onlinecite{purcell}
demonstrates this effect for singly-clamped multi-wall carbon
nanotubes. We expect that our predictions will soon be tested in
experiments on doubly-clamped SWNTs. 

We mention also one more paper published after the submission of our
manuscript, Ref. \onlinecite{Munich}, which shows measurements on
a suspended quantum dot. Though the focus of our study was on carbon
nanotubes, all the calculations can be immediately applied to this case
as well.

We thank Yu.~V.~Nazarov, P.~Jarillo-Herrero, L.~P.~Kouwenhoven and
C.~Dekker for discussions. This work was
supported by the Netherlands Foundation for Fundamental Research on
Matter (FOM) and ERATO. HSJvdZ was supported
by the Dutch Royal Academy of Arts and Sciences (KNAW).


\begin{thebibliography}{99}

\bibitem{Roukes1} For an overview, see M.~Roukes, Physics World {\bf
14}, 25 (2001).

\bibitem{Att} T.~D.~Stowe, K.~Yasumura, T.~W.~Kenny, D.~Botkin,
K.~Wago, and D.~Rugar, Appl. Phys. Lett. {\bf 71}, 288 (1997).

\bibitem{Cleland} A.N. Cleland and M.L. Roukes, Nature {\bf 392}, 160
(1998).

\bibitem{Roukes2} K.~Schwab, E.~A.~Henriksen, J. ~M.~Worlock, and
M.L. Roukes, Nature {\bf 404}, 974 (2000).

\bibitem{Casimir} H.~B.~Chan, V.~A.~Aksyuk, R.~N.~Kleiman,
D.~J.~Bishop, and F.~Capasso, Science {\bf 291}, 1941 (2001).

\bibitem{Roukes98} N.~F.~Schwabe, A.~N.~Cleland, M.~C.~Cross, and
M.~L.~Roukes, Phys. Rev. B {\bf 52}, 12911 (1995).

\bibitem{Levitov} A.~V.~Shytov, L.~S.~Levitov, and C.~W.~J.~Beenakker,
Phys. Rev.Lett. {\bf 88}, 228303 (2002).

\bibitem{Goth} L.~Y.~Gorelik, A.~Isacsson, M.~V.~Voinova, B.~Kasemo,
R.~I.~Shekhter, and M.~Jonson, Phys. Rev. Lett. {\bf 80}, 4526 (1998);
C.~Weiss and W.~Zwerger, Europhys. Lett. {\bf 47}, 97 (1999); T.~Nord,
L.~Y.~Gorelik, R.~I.~Shekhter, and M.~Jonson, Phys. Rev. B {\bf 65},
165312 (2002).

\bibitem{Erbe} M.~T.~Tuominen, R.~V.~Krotkov, and M.~L.~Breuer,
Phys. Rev. Lett. {\bf 83}, 3025 (1999); A.~Erbe, C.~Weiss, W.~Zwerger,
and R.~H.~Blick, Phys. Rev. Lett. {\bf 87}, 096106 (2001).

\bibitem{C60} H.~Park, J.~Park, A.~K.~L.~Lim, E.~H.~Anderson,
A.~P.~Alivisatos, and P.~L.~McEuen, Nature {\bf 407}, 57 (2000).

\bibitem{kim} P.~Kim and C.~M.~Lieber, Science {\bf 286}, 2148 (1999).

\bibitem{akita} S.~Akita, Y.~Nakayama, S.~Mizooka, Y.~Takano,
T.~Okawa, Y.~Miyatake, S.~Yamanaka, M.~Tsuji, and T.~Nosaka,
Appl. Phys. Lett. {\bf 79}, 1691 (2001).

\bibitem{rueckes} T.~Rueckes, K.~Kim, E.~Joselevich, G.~Y.~Tseng,
C.-L.~Cheung, and C.~M.~Lieber, Science {\bf 289}, 94 (2000).

\bibitem{sensors} See {\em e.g.} R.~H.~Baughman, C.~Cui,
A.~A.~Zakhidov, Z.~Iqbal, J.~N.~Barisci, G.~M.~Spinks, G.~G.~Wallace,
A.~Mazzoldi, D.~De Rossi, A.~G.~Rinzler, O.~Jaschinski, S.~Roth, and
M.~Kertesz, Science {\bf 284}, 1340 (1999).

\bibitem{dequesnes} M.~Desquesnes, S.~V.~Rotkin, and N.~R.~Aluru,
Nanotechnology {\bf 13}, 120 (2002).

\bibitem{kinaret} J.~M.~Kinaret, T.~Nord, and S.~Viefers,
Appl. Phys. Lett. {\bf 82}, 1287 (2003).

\bibitem{legoas} Q.~Zheng and Q.~Jiang, Phys Rev. Lett. {\bf 88},
045503 (2002); S.~B.~Legoas, V.~R.~Coluci, S.~F.~Braga, P.~Z.~Coura,
S.~O.~Dantas, and D.~S.~Galv\~{a}o, Phys Rev. Lett. {\bf 90},
055504 (2003).

\bibitem{Tombler} T.~W.~Tombler, C.~Zhou, L.~Alexseyev, J.~Kong,
H.~Dai, L.~Liu, C.~S.~Jayanthi, M.~Tang, and S.-Y.~Wu, Nature {\bf
405}, 769 (2000).

\bibitem{Cobden} J.~Nyg{\aa}rd and D.~H.~Cobden,
Appl. Phys. Lett. {\bf 79}, 4216 (2001).

\bibitem{Williams} P.~A.~Williams, S.~J.~Papadakis, M.~R.~Falvo,
A.~M.~Patel, M.~Sinclair, A.~Seeger, A.~Helser, R.~M.~Taylor~II,
S.~Washburn, and R.~Superfine, Appl. Phys. Lett. {\bf 80}, 2574
(2002). 

\bibitem{Franklin} N.~R.~Franklin, Q.~Wang, T.~W.~Tombler, A.~Javey,
M.~Shim, and H.~Dai, Appl. Phys. Lett. {\bf 81}, 913 (2002).

\bibitem{freq} B.~Reulet,  A.~Yu.~Kasumov, M.~Kociak, R.~Deblock,
I.~I.~Khodos, Yu.~B.~Gorbatov, V.~T.~Volkov, C.~Journet, and
H.~Bouchiat, Phys. Rev. Lett. {\bf 85}, 2829 (2000).

\bibitem{PKim} P.~Kim, L.~Shi, A.~Majumdar, and P.~L.~McEuen,
Phys. Rev. Lett. {\bf 87}, 215502 (2001).

\bibitem{GTKim} G.-T.~Kim, G.~Gu, U.~Waizmann, and S. ~Roth,
Appl. Phys. Lett. {\bf 80}, 1815 (2002). 

\bibitem{LL} L.~D.~Landau and E.~M.~Lifshits, {\em Theory of
Elasticity} (Pergamon, Oxford, 1986).

\bibitem{Armour1} S.~M.~Carr, W.~E.~Lawrence, and M.~N.~Wybourne,
Phys. Rev. B {\bf 64}, 220101 (2001) discuss a bistability as a result
of {\em externally applied} negative residual tension. We emphasize
that in our case it appears for an arbitrary residual tension, in the
regime when the charging effects drive the tube into the
strong-bending regime.

\bibitem{Munich} E.~M.~H\"ohberger, R.~H.~Blick, F.~W.~Beil,
W.~Wegscheider, M.~Bichler, and  and J.~P.~Kotthaus, Physica E {\bf
12}, 487 (2002); E.~M.~H\"ohberger {\em et al} (unpublished).

\bibitem{Yakobson} B.~I.~Yakobson, C.~J.~Brabec, and J.~Bernholc,
Phys. Rev. Lett. {\bf 76}, 2511 (1996).

\bibitem{purcell} S.~T.~Purcell, P.~Vincent, C.~Journet, and
Vu~Thien~Binh, Phys. Rev. Lett. {\bf 89}, 276103 (2002).

\end{thebibliography}
\end{document}